\documentclass{article}
\usepackage[T1]{fontenc}
\usepackage[utf8]{inputenc}
\usepackage{ismir} % Remove the "submission" option for camera-ready version
\usepackage{amsmath,cite,url}
\usepackage{graphicx}
\usepackage{multirow, booktabs, makecell}
\usepackage{xcolor}
\usepackage{subcaption}
\usepackage{soul}

\usepackage{float}
\usepackage{placeins}

\title{Learning to Predict Performance-induced \\ Emotion Differences
in Classical Piano Music}

\multauthor
   {Joann Ching$^1$ \hspace{1cm} Gerhard Widmer$^1$$^2$}
   {$^1$ Institute of Computational Perception, Johannes Kepler University Linz, Austria\\
   $^2$ LIT AI Lab, Linz Institute of Technology, Austria\\
   {\tt\small first.last@jku.at}
   }

\def\authorname{J. Ching, and G. Widmer}

\usepackage[bookmarks=false,pdfauthor={\authorname},pdfsubject={\pdfsubject},hidelinks]{hyperref}
\usepackage{cleveref}

\begin{document}

\maketitle

\begin{abstract}

Music is often used as a medium for communicating emotion, with performers shaping perceived affect through interpretation. This study addresses the challenge of identifying and predicting subtle changes in perceived emotion that are exclusively due to differences in performance.
We focus on classical solo piano music, using a set of 6 commercial recordings of Bach's Well-Tempered Clavier Book I, annotated in terms of valence and arousal.
By encoding the recordings through performance-specific features only, we isolate performance information from aspects of the composition itself, which tend to dominate the overall perceived emotional category. A preliminary analysis validates that these features vary meaningfully across performers. We then propose a relative regression framework, Delta-VA, to predict deviations in valence-arousal relative to an ``average'' performance, thereby focusing on the changes in emotion brought about by a specific way of playing a piece.
In addition to the standard $R^2$ regression score, we introduce geometric evaluation metrics to assess the preservation of pairwise differences between performances. Results indicate high directional consistency with the ground truth, but also a compression in prediction magnitude, indicating that the model tends to underestimate expressive performance effects.

\end{abstract}

\section{Introduction}\label{sec:introduction}

Music Emotion Recognition (MER) has been a continuing research topic in MIR, due to the central role of emotion and affect in music appreciation and the numerous application possibilities. The general approaches -- training classifiers on top of hand-selected audio features, or using deep neural networks directly on the recordings (e.g., \cite{cheuk2020triplet, malik2017mer, Dong2019BidirectionalCR}) -- have been rather indiscriminate with regard to the possible sources of emotion expression in music. 
Especially in Western classical music, \textit{performance} plays a crucial role in shaping or sharpening affective aspects of a piece. However, prior MER work has not generally sought to separate the performance component from the composition itself and to study its specific contribution to the affective experience.
As a result, emotion prediction models tend to rely on general characteristics of a piece, such as mode (e.g., the characterization of major pieces as "happy" and minor as "sad") or rather global characteristics such as tempo and loudness.
However, the same piece (composition) can be made to convey or evoke rather different affects or moods, if played differently. Performers employ complex timing, articulation, and phrasing strategies to shape their interpretation, and it has been shown, in controlled experimental contexts, that they can reliably communicate specific named emotions through their playing that listeners can then also quite reliably decode \cite{gabrielsson1996expression, akkermans2018decoding}.

It is these fine performance nuances and their contribution to perceived emotional character that we wish to investigate here. Specifically, we use a collection of recordings, by six renowned pianists, of classical piano music (Bach's Well-Tempered Clavier) annotated in terms of the emotion dimensions valence and arousal. To disentangle the subtle effects of performance from the broader, and perhaps stronger, effects of the composition itself, we set up a difference prediction scenario in the following way: (a) we represent the recordings in terms of a set of features (a \textit{performance codec} \cite{cancino2018computational}) that capture specifically, and exclusively, performance-related aspects (expressive timing, dynamics, articulation), but no music-structural features such as key or mode that would characterize the piece itself; and (b) instead of attempting to predict absolute position in valence-arousal space (which is likely dominated by the character of the piece
), we train models to predict the deviation of a performance from a hypothetical ``average'' performance. 
We first conduct exploratory analyses to establish the reliability of the underlying methods (music transcription, on which the calculation of performance features relies; and the expressivity of the actual performance encoding). 
We then evaluate baseline regression models that predict emotion directly from the performance codec, and show that our difference prediction model, Delta-VA, is more appropriate, with our performance features providing a meaningful signal for modeling differences induced through expressive interpretation. In particular, an evaluation of the model on the task of comparing the emotional differences between pairs of alternative recordings suggests that this kind of model may support tasks such as classical music performance recommendation.

\section{Related Work}\label{sec:related}

Although performance-based features have not been widely used for emotion modeling, they have been extensively studied in research on computational performance modeling (e.g., \cite{Grachten01122012,cancino2017evaluation}).
Similarly, expressive interpretation across performers has been investigated, e.g., in the Mazurka Project \cite{shi2021computational}.
Performance features are also commonly employed in tasks such as expressive performance rendering, where models aim to generate human-like musical interpretations from score-based inputs \cite{cancino2018computational, fujishima2019rendering, jeong2019virtuosonet, borovik2023scoreperformer, zhang2024dexter}. Despite their widespread use in modeling expressive variations, their role in understanding perceived affect in performances remains underexplored. This motivates our investigation into utilizing such features as a basis for modeling emotional differences across interpretations.

To understand how a performer's interpretation affects perceptual phenomena, previous studies have compared professional renditions against deadpan performances (e.g., \cite{gabrielsson1999performance}).
The \textit{Well-tempered Clavier (WTC)} by J.S. Bach has been proposed by several authors as an ideal choice for the study of naturalistic communication of emotion, 
due to its interpretive openness. The absence of composer-specified tempi and expression markings, along with relatively ambiguous emotional cues in the score, allows for substantial variation across performances. For example, Battcock and Schutz\cite{battcock&schutz} (referred to as B\&S hereafter) examined possible sources of emotional expression in the collection by considering several musical predictors, such as attack rate, mode,
and pitch, and found -- not surprisingly -- that attack rate (which is related both to the note density of the composition and to performance tempo) is the strongest predictor of arousal, and mode is closely tied to perceived valence. They further investigated the weight of composer-controlled versus performer-manipulated cues in a later study with an expanded collection of pianists\cite{Battcock2021IndividualizedIE}, and observed a higher contribution in performer-manipulated cues.
The most recent in literature, Anderson et al.\cite{anderson2025beyond} looked into the contribution of composition versus performance features by comparing differences in emotional ratings of recordings from a Grammy-winning pianist against deadpan versions. Their result further strengthens previous findings, namely, that perceived valence is strongly based on compositional cues, while performance cues contribute more to arousal.

In line with \cite{battcock&schutz}, %, Battock2021IndividualizedIE} 
Chowdhury et al.\cite{chowdhury2021midlevel} introduced an extended WTC dataset, which comprises recordings by six famous pianists
of the entire WTC Book I, which is also used in the present study. 
The data collection procedure followed that of B\&S.
While the focus in \cite{chowdhury2021midlevel} is also on performance-wise variations, their main contribution is further evidence for the relevance of so-called \textit{mid-level perceptual features} (which are themselves learned from audio) to emotion prediction. In line with prior work, they identified features related to modality as predictive of valence, and rhythm- and articulation-related properties (including perceived ``melodiousness'') as related to arousal.
But again, this investigation remained at a rather abstract level: the mid-level features may be partly related to performance aspects, but it is unclear to what extent. Obtaining precise performance details would have required automatic transcription from the audio recordings, which was considered too unreliable at the time. In our present study, we will, for the first time, use the latest state-of-the-art transcription models to automatically obtain precise performance features for emotion prediction from commercial recordings.

\section{Data Representation}\label{sec:data_rep}

As discussed above, when considering feature-based
representations of music,
we distinguish between \textit{score features} and \textit{performance features}, where the former relate to aspects of the piece, the composition itself (e.g., pitches, meter, mode, key), and the latter capture aspects shaped by the performer's interpretation (e.g., precise timing and dynamics).
Given a dataset with multiple performances of the same piece and our goal of modeling performance-induced emotion differences, 
we focus exclusively on performance features to isolate interpretive effects. Each performance is encoded using a set of four performance-related features, referred to as the \textit{performance codec}, following \cite{cancino2018computational}.\footnote{Complementary code is provided on \url{github.com/joann8512/performance-rlt}}

\subsection{Dataset}\label{subsec:dataset}
The CP-WTC Dataset\footnote{\label{fn:zenodo}\url{https://doi.org/10.5281/zenodo.21693024}}, introduced in \cite{chowdhury2021midlevel}, comprises recordings of J.S.Bach's Well-Tempered Clavier, Book I, performed by six pianists: Glenn Gould, Friedrich Gulda, Angela Hewitt, Sviatoslav Richter, András Schiff, and Rosalyn Tureck. The book includes 48 pieces in total (24 preludes and 24 fugues, covering all keys). Following the strategy of B\&S\cite{battcock&schutz}, the first eight bars of each recording are annotated with valence and arousal values, resulting in 288 segments with the six performers. %

As the task focuses on comparing performances of the same piece, the data are split by performer rather than randomly.
Given $N = 6$ pianists, we adopt an exhaustive cross-validation scheme over performer-wise splits: $N_{train}=4$ are used for training, with the remaining two assigned to validation and test. The total number of folds is given by:
    $\binom{N}{N_{\mathit{train}}} \times 2 = 30$.

\subsection{Performance Codec}\label{subsec:codec}

\begin{figure}[t]
 \centering
  \includegraphics[width=\linewidth]{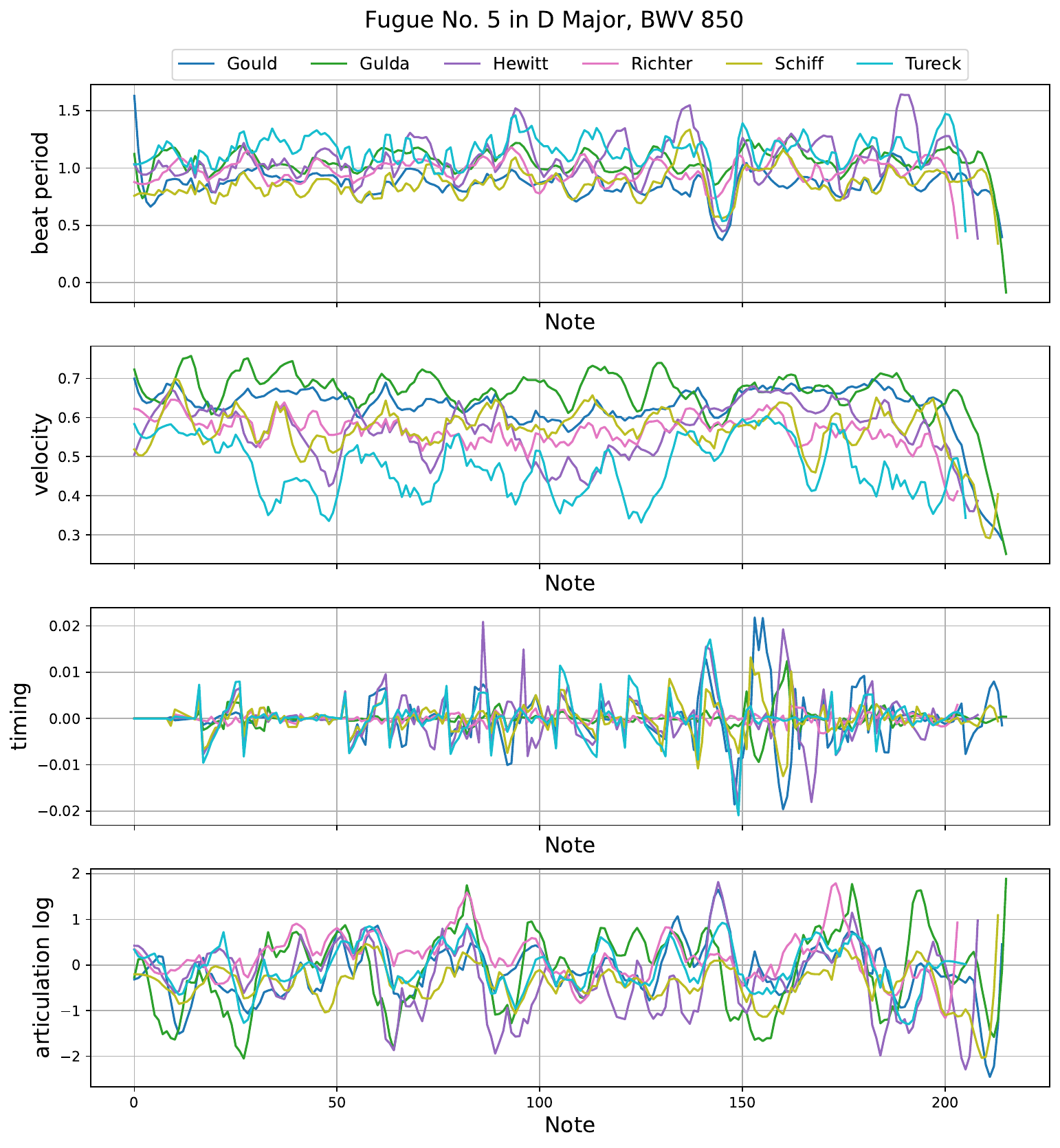}
 \caption{Performance codec representation for Fugue No. 5 in D Major: the execution generally follows the same trend as the underlying musical content is the same, while variations on the same note are present.} 
 \label{fig:codec_ex}
\end{figure}

The \textit{Performance Codec} is a set of four parameters, originally proposed and formally defined in the expressive rendering framework \textit{Basis Mixer}\cite{cancino2018computational}; these parameters are computed for each note ($n_i$) in the score (Figure \ref{fig:codec_ex}). They capture expressive controls relative to the score, rather than absolute compositional content (refer to \cite{cancino2018computational} for the computation formulae). The four parameters are:
\begin{itemize}
  \item \textbf{Beat Period}. This parameter represents local tempo, defined as the ratio between the inter-onset intervals (IOI) of consecutive notes in the performance and in the score.% 

  \item \textbf{Velocity}. In the piano, perceived loudness is a subjective measure related to hammer velocity, which is represented as MIDI velocity values from 0 to 127. Although the relationship between loudness and MIDI velocity is not strictly linear\cite{Goebl2003MeasurementAR},
  normalized MIDI velocity is used as a proxy for dynamics.

  \item \textbf{Timing}. This parameter captures temporal deviations from the nominal score timing, including expressive phenomena such as chord spread\cite{fu2015timing} and melody lead\cite{Goebl2001MelodyLI}. Positive values indicate anticipation, negative values indicate delayed onset.
  The timing of a note  
  is computed by first estimating the average performed onset of all notes sharing the same score onset, and then measuring the deviation from the actual performed onset.
  
  \item \textbf{Articulation ratio} Articulation denotes the relative duration of a performed note compared to its notated duration, capturing expressive variations such as \textit{legato} and \textit{staccato}. 
\end{itemize}

Together, these features compose the \textit{performance codec}. Note that, in combination with the score of the piece, this yields a \textit{lossless} representation of a performance:\footnote{The \textit{pedaling} dimension is excluded from the codec due to the challenges of modelling it objectively, as the mapping of CC-64 values to acoustic effects is specific to each piano.} given the score and a performance encoding in the form of these four note-by-note feature sequences, the performance can be unambiguously reconstructed.

\subsection{Obtaining Performance Details via Transcription}
Performance features are derived with Partitura\cite{partitura_mec} from a symbolic (MIDI) representation. As our professional performances are given as audio recordings, MIDI is obtained via a pretrained transcription model. However, prior work has shown that such models are sensitive to recording conditions, particularly for velocity estimation \cite{edwards2024datadriven, martak2026trans}, which is critical here given the variability in noise and reverberation in commercial recordings. We therefore first systematically evaluate transcription robustness under controlled perturbations. Using the ASAP dataset\cite{foscarin2020asap}, which provides aligned audio-MIDI pairs, we generate three audio variants: \textit{noise}, \textit{reverb}, and \textit{noise+reverb}. Reverberation is applied using an impulse response with a 2-second delay, while additive white noise is introduced at a level of 0.01. And we evaluate four recent piano transcription models: Kong et al.\cite{kong2020HighRes}; Edwards et al.\cite{edwards2024datadriven}; Toyama et al.~(hFT)\cite{toyama2023}, and Yan et al. (Transkun)\cite{yan2024scoring}, with emphasis on the robustness of velocity estimation. 

Transcription performance is evaluated using \textit{mir\_eval} \cite{Raffel2014mir_eval} and \textit{mpteval} \cite{hu2024towards}. 
Rather than absolute accuracy, we prioritize correlation-based metrics, 
as relative loudness variation is more important for describing expressive dynamics. 
\textit{mir\_eval}, widely used for transcription evaluation, reports metrics for velocity-only and note+velocity, with the latter computed only for correctly matched notes based on pitch and onset/offset timing. For velocity-only, we include F1 (detection accuracy under a velocity tolerance), correlation (linear agreement between predicted and ground-truth velocities), and MAE (absolute error). 
The dynamic score from \textit{mpteval} measures correlation between ground-truth and estimated performances through the loudness ratio between melody and bass lines, serving as a proxy for how well dynamic relationships are preserved.
For space reasons, we omit the detailed comparative results table here. Ultimately, \textbf{Transkun} comes out as the most robust model, topping its competitors in almost all measures, 
on our augmented data. 
Notably, the model applies input normalization, which mitigates loudness differences across recordings.

\section{Analysis of Feature Variability}

As illustrated in Figure \ref{fig:codec_ex}, different performances of the same piece exhibit similar structural patterns, as they share the same underlying composition. To ensure that the performance-based features capture meaningful variation across performers rather than mainly encoding piece identity, we conduct the following analyses.

\begin{figure}[h]
\centering

\begin{subfigure}[]{\linewidth}
    \centering
    \includegraphics[width=0.9\linewidth]{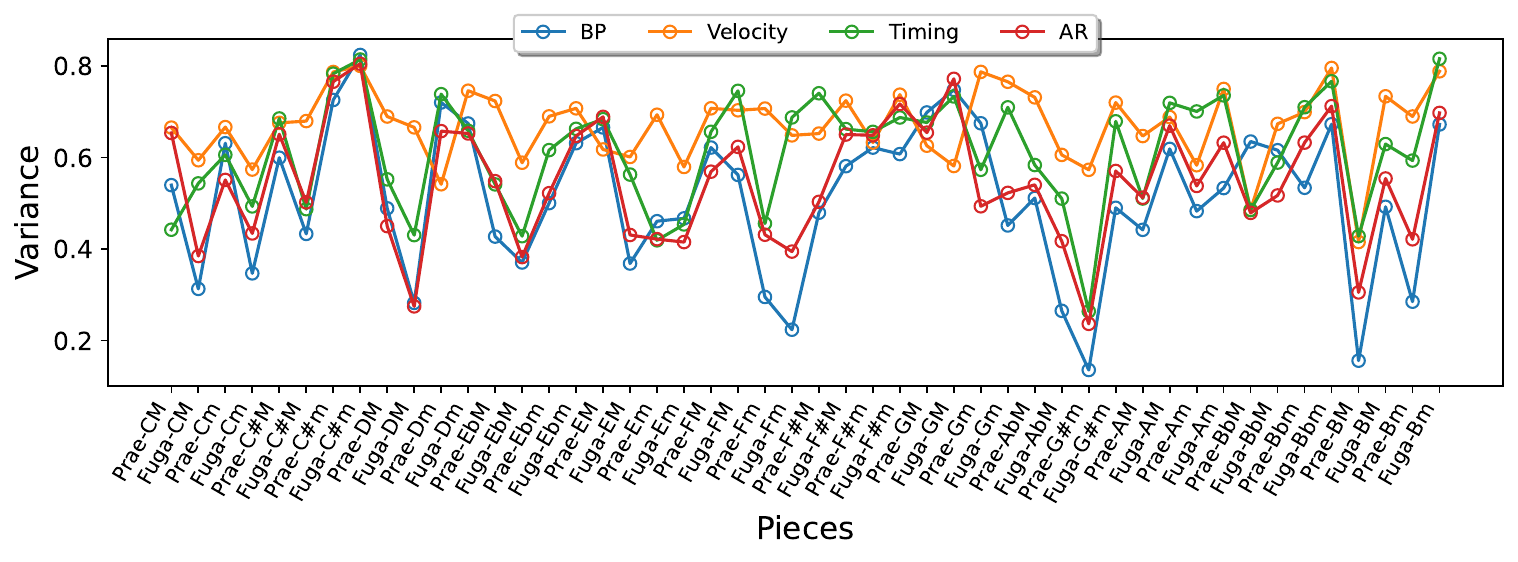}
    \caption{Per-piece variance of each feature across all performers, illustrating the effect of musical content on expressive variability.}
    \label{fig:feat_piece}
\end{subfigure}
\par\bigskip
\begin{subfigure}[t]{0.46\linewidth}
    \centering
    \includegraphics[width=\linewidth]{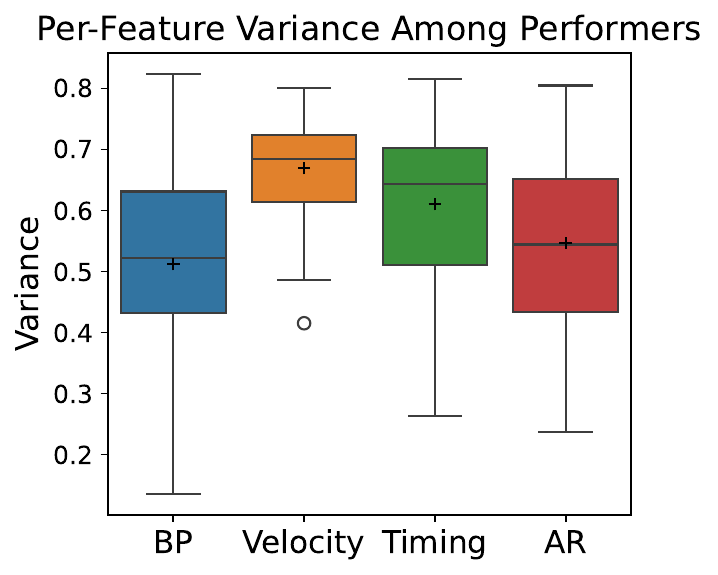}
    \caption{Boxplot view of the variances per-feature averaged across all 48 pieces.}
    \label{fig:variance}
\end{subfigure}
\hfill
\begin{subfigure}[t]{0.46\linewidth}
    \centering
    \includegraphics[width=\linewidth]{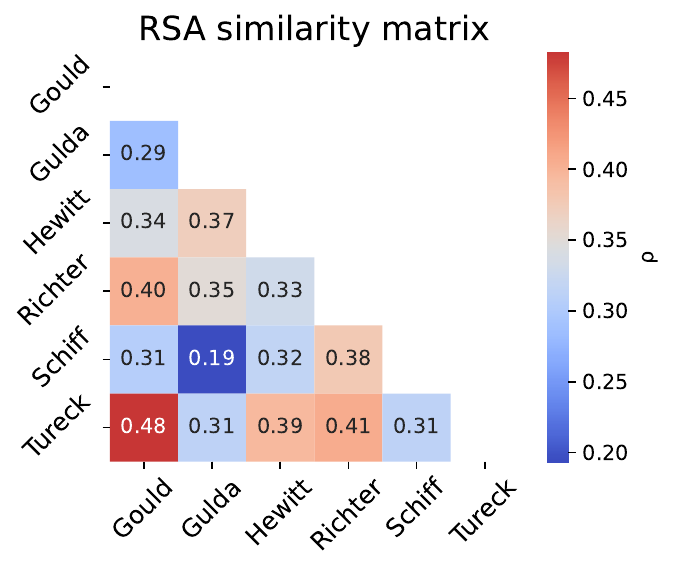}
    \caption{RSA matrix across performer pairs. Lower values indicate greater dissimilarity.}
    \label{fig:rsa}
\end{subfigure}

\caption{Analysis of performance variability.}
\label{fig:combined_analysis}
\end{figure}

Feature-level variability is quantified by computing variance across the six performances of each piece. 
As Fig.\ref{fig:variance} shows, velocity exhibits the highest and most consistent variance,  
indicating greater expressive freedom. In contrast, timing, articulation, and beat period show lower average variance with a wider spread, suggesting stronger dependence on musical content -- some pieces permit more interpretive flexibility than others (supported by Fig.~\ref{fig:feat_piece}).

To further examine whether the representation captures performer-specific variation rather than merely piece identity, 
we employ Representational Similarity Analysis (RSA)\cite{kriegeskorte2008rsa}.%
Treating each piece as a stimulus and each performer's feature sequence as its representation, we compare the performances via Spearman correlation between pairwise representational distance matrices (the RDMs). 
The resulting similarities (Figure \ref{fig:rsa}) range from 0.19 to 0.48, indicating moderate agreement in structural relationships across pieces. While some performer pairs exhibit higher similarity, it suggests a balance between shared structure and distinct expressive characteristics overall.

\section{Learning from Performance}\label{sec:methodology}

We now turn to our main goal -- predicting perceived emotion from performance details -- and compare two formulations: (i) standard regression models that predict absolute valence–arousal values (baselines), and (ii) a relative regression formulation that predicts deviations induced by performance variations. This distinction aligns with our earlier discussion by explicitly modeling performance-induced variation, rather than treating each performance as a fixed point in the valence–arousal space.

To accommodate pieces of different lengths, models are trained on segments of N notes where shorter segments are padded.
As mentioned before, the length of the performance codecs is the number of notes; the 8 bars of annotated segments have varying lengths. We perform cyclic padding, 
such that padding cannot be read as ``information'' that obscures the actual performance content. In our experiments, we take the maximum length where N = 314.

Although the objectives differ, both formulations produce predictions in the same valence–arousal space, enabling direct comparison.
We evaluate all models using Mean Squared Error (MSE) (averaged over the two dimensions) and the coefficient of determination ($R^2$-score), which indicates how well the regression model fits the observed data, enabling consistent comparison across approaches.
Baselines include regression models of increasing complexity that predict absolute values, along with a rank-based model
adapted from \cite{zha2023rnc}, which incorporates relative ordering information while still producing absolute predictions.

\subsection{Baseline Regressors} \label{subsec:baseline}
To progressively assess the informativeness of the performance-based features to predict emotion, three models are trained: (1) a linear regressor, (2) a k-Nearest Neighbor (kNN) regressor, and (3) a multi-layer perceptron (MLP). For the first two models, each performance is represented using a 24-dimensional descriptor vector, obtained by summarizing the four feature types (velocity, timing, articulation, beat period) from the performance codec via mean, standard deviation, minimum, maximum, range ($\mathit{max-min}$), and slope. This compact representation captures broad trends in the performance features while reducing dimensionality. The MLP, in contrast, operates on the original (non-summarized) performance codec and consists of four fully-connected layers with ReLU activation applied between layers, progressively reducing the input to a two-dimensional valence-arousal prediction. The experimental results (Table \ref{tab:results}) show improvements over both linear and kNN regressors.

\subsection{Rank-based Regression}\label{subsec:rank}
As noted in \cite{zha2023rnc}, end-to-end regression models  
may learn representations that fail to preserve the continuous relationships underlying the task, leading to suboptimal results. 
To address this, we adapt the RnC loss from \cite{zha2023rnc} to the valence-arousal space. Unlike the scalar setting considered in \cite{zha2023rnc}, this space is two-dimensional, spanned by two largely independent factors, and does not have a unique ordering of samples. Therefore, we employ two separate embedding branches for valence and arousal. The model first produces a global embedding ($y_{emb}$), which is optimized with the RnC loss to enforce relative ordering between samples. This embedding is then projected through two separate linear layers to obtain dimension-specific predictions, $y_{\mathit{valence}}$ and $y_{\mathit{arousal}}$, with additional L2 supervision applied to each. By modeling relative relationships rather than treating samples independently, this approach yields a substantial improvement over the simpler baseline regressors (Table \ref{tab:results}).

\subsection{Relative Prediction}\label{subsec:delta_pred}

\begin{figure}[h]
\centering
    \includegraphics[width=0.8\linewidth]{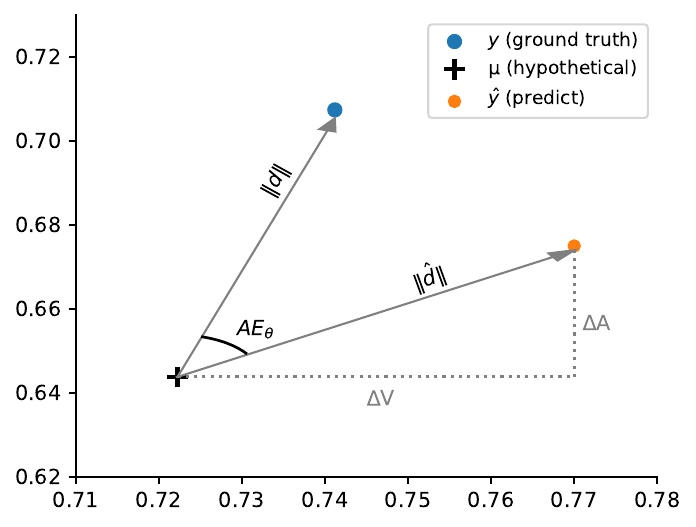}
\caption{Conceptual sketch of Delta-VA on one example. The model is trained to predict deviations $\Delta{V}$ and $\Delta{A}$ by minimizing MSE. An ideal prediction would result in an angular error of 0.} 
\label{fig:obj}
\end{figure}

As prior work suggests, overall perceived valence and arousal depend heavily on piece-specific aspects such as pitch, rhythm, and mode. The performance codec, however, encodes only expressive aspects of the performance, giving a representation in which score-related information is largely absent, or only implicitly present. Models relying only on performance features can thus not be expected to perform well in predicting the absolute positioning of a piece and its performances in valence-arousal space.
Our hypothesis here is, however, that performance details should permit us to predict \textit{differences} in perceived affect induced by interpretation. Different performances of the same piece are expected to occupy a relatively compact region of the valence-arousal space, the placement of which is determined by the piece's characteristics. Therefore, we reformulate the performance-to-emotion regression task as a relative difference prediction problem, which we call \textit{Delta-VA}.

For a given performance (recording), the model predicts deviations ($\Delta{V}$ and $\Delta{A}$) relative to a piece-specific reference point in the valence-arousal space, which we refer to as the \textit{score-state} ($\mu_{score}$). The score-state represents the average affective annotation for a given piece and is computed as the centroid of all annotations of the piece. The predicted outputs thus model how a given performance deviates from this reference state, capturing performance-induced variations in perceived emotion.

If we define our prediction target $y_{\mathit{dev}}$ (deviation from score-state in the valence and arousal dimensions) as 
\begin{equation}
y_{dev} =(\Delta{Valence}, \Delta{Arousal}),
\end{equation}
so that the actual absolute valence-arousal position $\hat{y}$ of the recording would be
\begin{equation}
\hat{y}=\mu_{score}+{y}_{dev},
\end{equation}
where $\mu_{score}$ is derived from the training corpus.
Then we can define our loss function as
\begin{equation}
    MSE(y, \hat{y}).
\end{equation}
The model is then trained by minimizing the mean squared error (again, averaged over the two dimensions) between $y$ (ground truth) and $\hat{y}$ (predicted).
In this way, the final emotion predictions of the model are of the same kind as, and directly comparable to, the absolute valence-arousal predictions made by the baseline regression models.

As the performance codec covers four feature types of differing ranges and semantics, each feature is processed independently through a dedicated linear projection layer. Specifically, each 314-dimensional feature block is mapped to a 16-dimensional embedding, allowing the model to learn feature-specific representations. The resulting embeddings are then concatenated to form a joint representation, which is passed through a shared fully connected layer with ReLU activation to integrate information across feature types. Finally, two separate linear output heads are used to predict valence and arousal, respectively.

\begin{table}[t!]
\centering
\begin{tabular}{r|cc}
\toprule
Model & \textbf{MSE} & \boldmath{$R^2$} \\ \Xhline{1.5pt}
Linear & \begin{tabular}[c]{@{}c@{}}.183\\(.152 / .214)\end{tabular} 
       & \begin{tabular}[c]{@{}c@{}}.196\\(.346 / .046)\end{tabular} \\
\hline
kNN & \begin{tabular}[c]{@{}c@{}}.160\\(.141 / .179)\end{tabular} 
    & \begin{tabular}[c]{@{}c@{}}.375\\(.452 / .298)\end{tabular} \\
\hline
MLP & \begin{tabular}[c]{@{}c@{}}.156\\(.129 / .183)\end{tabular} 
    & \begin{tabular}[c]{@{}c@{}}.410\\(.450 / .371)\end{tabular} \\
\hline
Rank-based & \begin{tabular}[c]{@{}c@{}}.072\\(.065 / .079)\end{tabular} 
           & \begin{tabular}[c]{@{}c@{}}.726\\(.721 / .732)\end{tabular} \\
\Xhline{1.2pt}
Delta-VA & \begin{tabular}[c]{@{}c@{}}\textbf{.014}\\(.017 / .010)\end{tabular} 
         & \begin{tabular}[c]{@{}c@{}}\textbf{.786}\\(.708 / .864)\end{tabular} \\
\bottomrule
\end{tabular}
\caption{Results of all models experimented with. All are evaluated by mean squared error and $R^2$ score. In parentheses are the scores for arousal and valence individually.
}
\label{tab:results}
\end{table}

\section{Analysis}

\begin{figure*}[!t]
\centering
\begin{subfigure}[t]{0.24\linewidth}
    \centering
    \includegraphics[width=\linewidth]{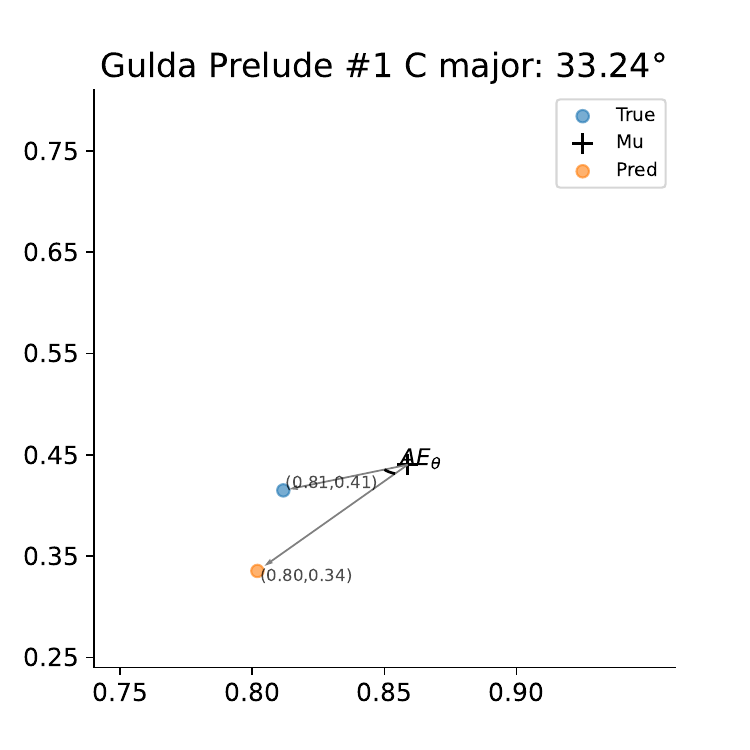}
\end{subfigure}
\begin{subfigure}[t]{0.24\linewidth}
    \centering
    \includegraphics[width=\linewidth]{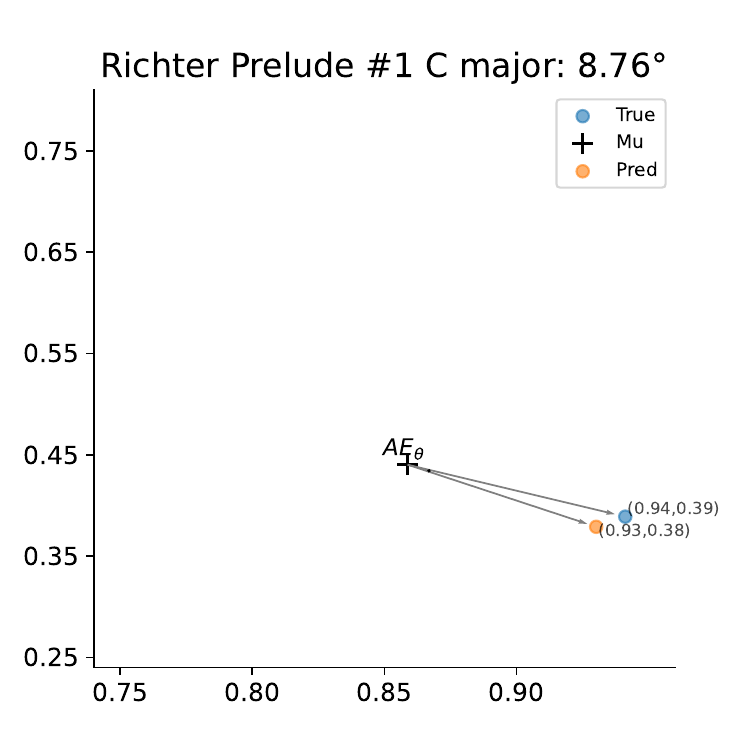}
\end{subfigure}
\begin{subfigure}[t]{0.24\linewidth}
    \centering
    \includegraphics[width=\linewidth]{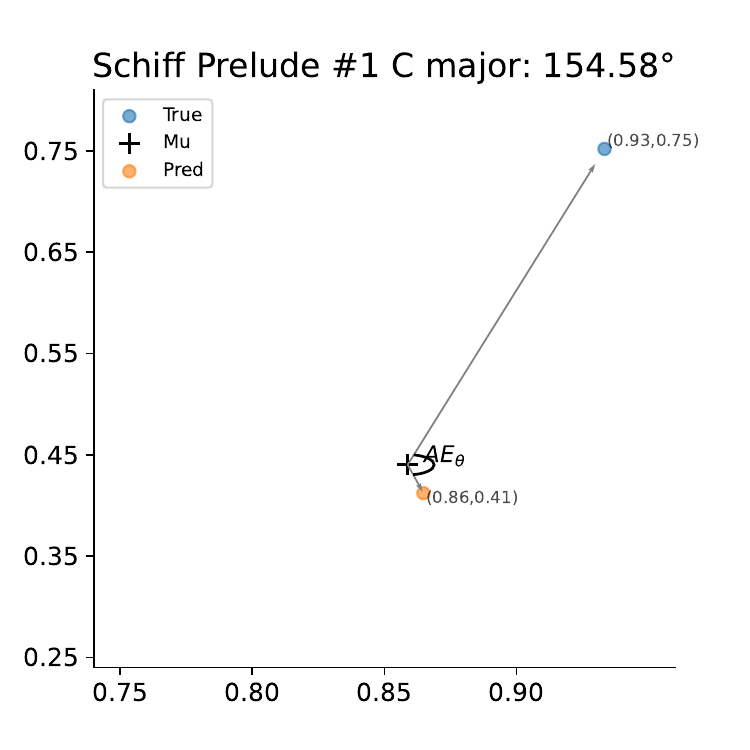}
\end{subfigure}
\begin{subfigure}[t]{0.24\linewidth}
    \centering
    \includegraphics[width=\linewidth]{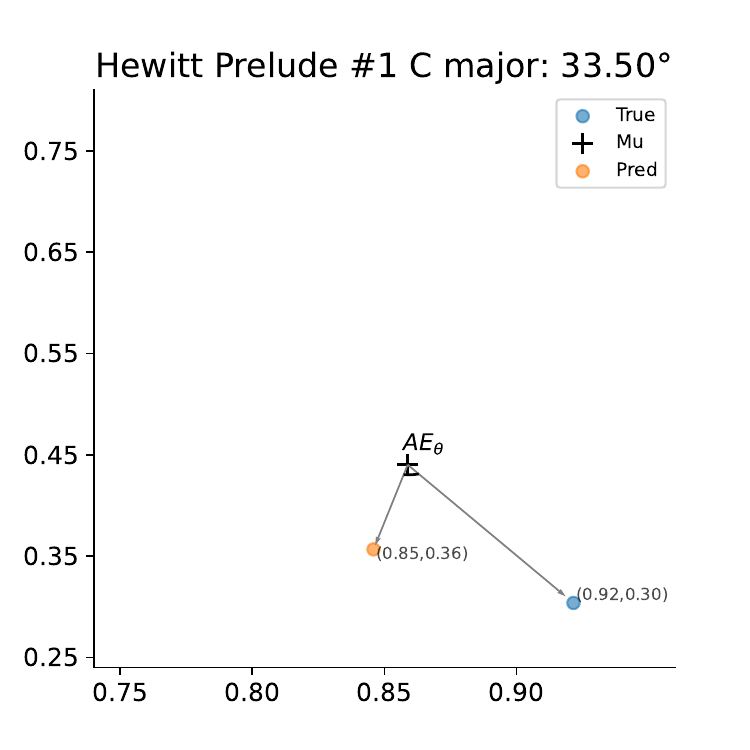}
\end{subfigure}

\begin{subfigure}[t]{0.24\linewidth}
    \centering
    \includegraphics[width=\linewidth]{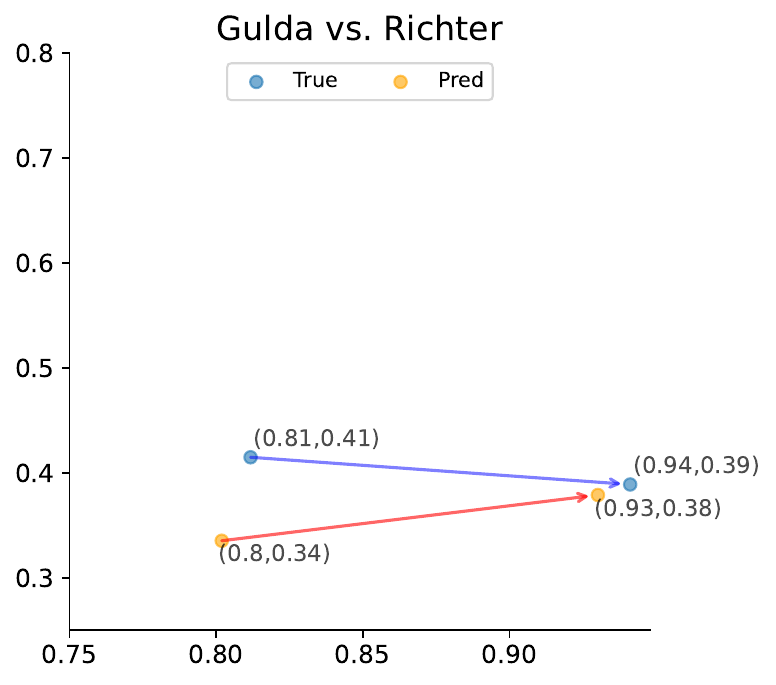}
\end{subfigure}
\begin{subfigure}[t]{0.24\linewidth}
    \centering
    \includegraphics[width=\linewidth]{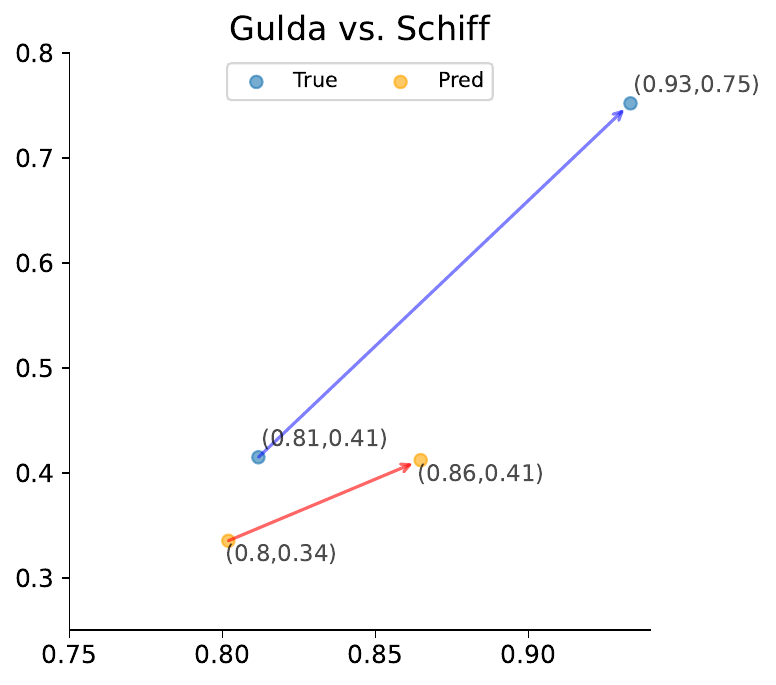}
\end{subfigure}
\begin{subfigure}[t]{0.24\linewidth}
    \centering
    \includegraphics[width=\linewidth]{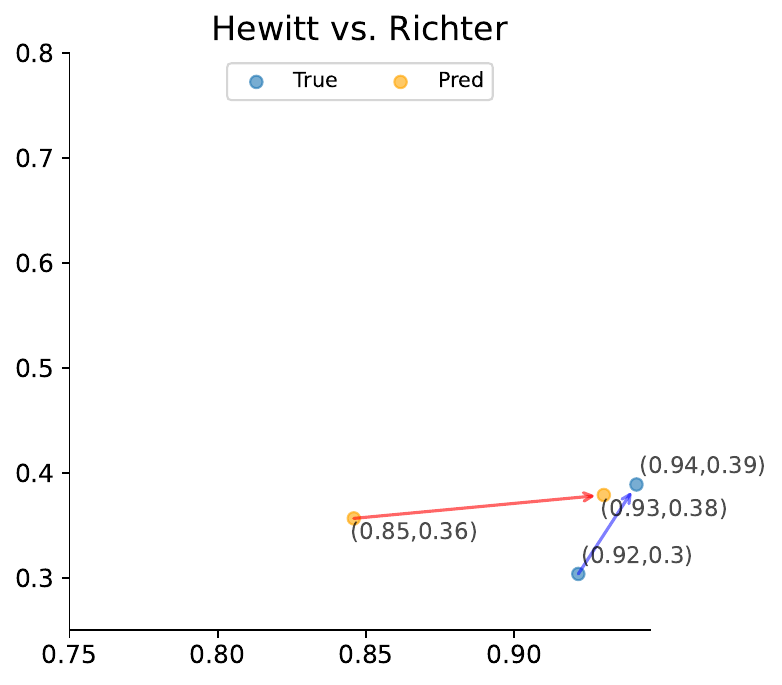}
\end{subfigure}
\begin{subfigure}[t]{0.24\linewidth}
    \centering
    \includegraphics[width=\linewidth]{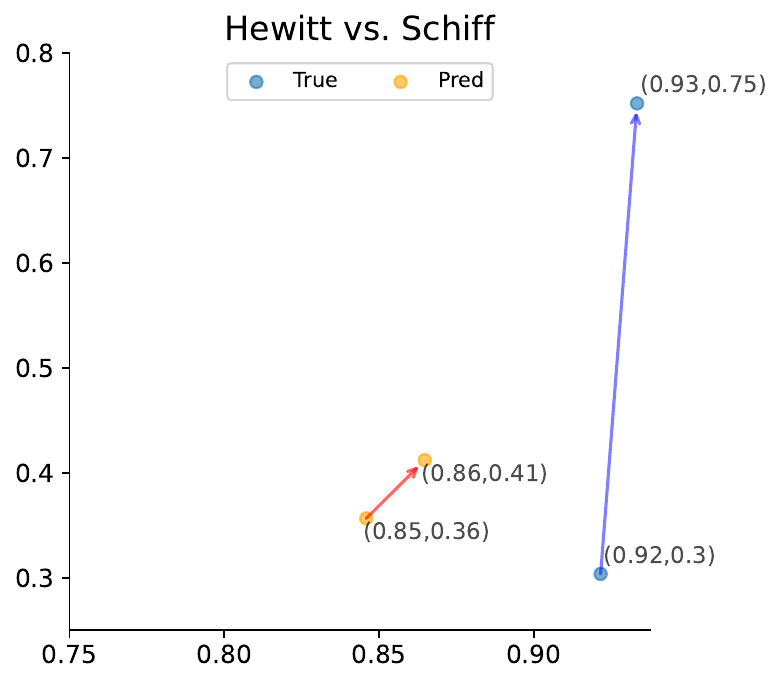}
\end{subfigure}
\caption{Prediction results of Delta-VA under the training objective view (top) and the alternative view (bottom). The top row shows four examples with prediction and ground truth; gray arrows indicate the vectors used to compute angular error. The bottom row shows four (out of the six possible) pairwise comparisons involving the same pianists and the same piece,
but with vectors connecting performance pairs (true vs predicted), illustrating how well differences are preserved.}

\label{fig:result}
\end{figure*}

As mentioned in Section \ref{sec:methodology}, all models -- including \textit{Delta-VA}, are evaluated by MSE and $R^2$-score (Table \ref{tab:results}). 
However, simply assessing the performance by a high $R^2$-score can be misleading, especially in our setting of predicting performance-induced deviations. 
While it indicates that the predictions are numerically close to the ground truth, it does not capture whether the relational structure between performances is preserved. 
To address this limitation, we introduce an additional geometric score, 
the \textit{Mean Angular Error}, which quantifies how well the predicted direction aligns with the ground truth with respect to the score-state.  

Let \textit{a} and \textit{b} be two performances with annotations $y_a$=($V_a$, $A_a$) and $y_b$=($V_b$, $A_b$). Then 
$d_{ab}$ and $\hat{d}_{ab}$
denote the ground-truth and predicted affective shift vectors, respectively. \textbf{Mean Angular Error} (MAE$_\theta$) measures the angular gap between $\hat{d}$ and $d$, capturing directional alignment averaged over $N$ pairs of performances:

\begin{equation}
    \text{MAE}_\theta = \frac{1}{N} \sum_{i=1}^{N} \arccos \left( 
    \frac{\hat{d}_{ab} \cdot d_{ab}}
    {\|\hat{d}_{ab}\| \, \|d_{ab}\|}
    \right).
\end{equation}

To ensure comparability with baseline models, MAE$_\theta$ is not included in the training objective $\mathcal{L}$. As a result, although the proposed model achieves an $R^2$-score of 0.786, it exhibits a relatively high mean angular error of 88.28$^{\circ}$. 
Importantly, this evaluation remains tied to the original objective, where deviations are defined with respect to the score-state. 
However, since the model predicts relative deviations, evaluating them against a fixed reference does not capture whether the relational structure between performances is preserved. 
We therefore adopt an alternative evaluation perspective, which assesses differences \textit{between} (pairs of) different performances rather than deviations from the score-state.

This view is particularly relevant in application scenarios such as classical music performance recommendation. For example, a query such as ``give me a performance that is more positive and less aggressive than X" requires recognizing differences between performances, rather than deviations from an imagined or hypothetical reference (e.g., the score-state). In such cases, it is not necessary for the model to infer absolute positions in valence-arousal space; instead, the \textit{direction} of change is what matters. 
To evaluate this, we consider pairs of performances and construct difference vectors between them for both ground-truth and predictions. The angular difference is then computed between these vectors. In this setting, a small angle indicates that the model successfully captures the relative direction of change between performances, independent of any global reference frame (the score-state). We further introduce an additional metric, the \textit{Mean Magnitude Ratio} (MMR), to assess how well predicted magnitude differences are preserved, as well as the systematic under- or over-estimation of affective shift intensity, defined as:
    \begin{equation}
         \text{MMR} = \frac{1}{N} \sum_{i=1}^{N} 
         \frac{\|\hat{d}_{ab}\|}{\|d_{ab}\|}.
    \end{equation}

Under this view, the model achieves a mean angular error of 8.374$^{\circ}$ $\pm$ 10.552 (with 0$^{\circ}$ indicating perfect directional alignment) and a mean magnitude ratio of 0.478 $\pm$ 0.184 (optimal at 1.0). While the predicted ($\Delta{Valence}$, $\Delta{Arousal}$) vectors exhibit a degree of magnitude compression, the directional component is largely preserved, as reflected by the low angular error.
For illustration, four performances of the Prelude \#1 in C major are taken as an example in Figure \ref{fig:result}. The top row graphically evaluates the predictions in relation to the hypothetical score-state, conforming to the training objective; the bottom row adopts the alternative view of between-performance difference prediction.
Taking the prediction result of Schiff, for example, although the angular error is large under the score-state reference (plot 3 in the upper row), pairwise comparisons involving Schiff (2 and 4 in the lower row) reveal that the direction of difference in valence and arousal is largely preserved.

\section{Conclusion}\label{section:conclusion}
In this work, we investigated Music Emotion Recognition utilizing performance-related features alone. 
In particular, we proposed the Delta-VA framework to learn from performance features to predict deviations in perceived valence and arousal that are an effect of differences in expressive interpretation. Through variance analyses and representational similarity comparisons, we confirmed that performance features exhibit meaningful variations across performers, providing a reliable signal for modeling relative affective shifts. Our results demonstrate success in modeling differences induced through expressive interpretation, rather than absolute positions, capturing meaningful structure in perceived emotion. In particular, the model performs well in capturing the direction (if not the extent) in which one performance affectively differs from another. We propose this framework as a step toward performance-centered modeling of musical emotion, highlighting the role of expressive variability as a key factor in perceived affect, and supporting tasks such as classical music performance recommendation.

\section{Acknowledgments}

% You may include an optional Acknowledgments section in your camera-ready version to refer to any individuals or organizations that should be acknowledged in your paper. \textbf{Do not include the Acknowledgments section in your submitted manuscript.} The Acknowledgments section does \textit{not} count towards the page limit for scientific content. 
This work is supported by the European Research Council (ERC) under the EU’s Horizon 2020 research \& innovation programme, grant agreement No. 101019375 (Whither Music?).

% For BibTeX users:
\bibliography{ISMIRtemplate}

% For non BibTeX users:
%\begin{thebibliography}{citations}
% \bibitem{Author:17}
% E.~Author and B.~Authour, ``The title of the conference paper,'' in {\em Proc.
% of the Int. Society for Music Information Retrieval Conf.}, (Suzhou, China),
% pp.~111--117, 2017.
%
% \bibitem{Someone:10}
% A.~Someone, B.~Someone, and C.~Someone, ``The title of the journal paper,''
%  {\em Journal of New Music Research}, vol.~A, pp.~111--222, September 2010.
%
% \bibitem{Person:20}
% O.~Person, {\em Title of the Book}.
% \newblock Montr\'{e}al, Canada: McGill-Queen's University Press, 2021.
%
% \bibitem{Person:09}
% F.~Person and S.~Person, ``Title of a chapter this book,'' in {\em A Book
% Containing Delightful Chapters} (A.~G. Editor, ed.), pp.~58--102, Tokyo,
% Japan: The Publisher, 2009.
%
%\end{thebibliography}

\end{document}